\documentstyle[emulateapj] {article}

\lefthead{EVANS ET AL.}
\righthead{CO IN IR-EXCESS QSOs}

\begin{document}

\title{Molecular Gas in Infrared-Excess, Optically-Selected QSOs and
the Connection with Infrared Luminous Galaxies}

\author{A. S. Evans\altaffilmark{1},
D. T. Frayer\altaffilmark{2}
J. A. Surace\altaffilmark{2},
\& D. B. Sanders\altaffilmark{3}}

\altaffiltext{1}{Department of Physics \& Astronomy,
SUNY, Stony Brook, NY, 11794-3800: aevans@mail.astro.sunysb.edu}
\altaffiltext{2}{SIRTF Science Center, California
Institute of Technology, MS 314-6, Pasadena, CA 91125}
\altaffiltext{3}{Institute for Astronomy, University of Hawaii, 2680
Woodlawn Dr., Honolulu, HI 96822}

\begin{abstract}

The initial results of a millimeter (CO) survey of infrared-excess,
optically-selected quasars from the Palomar-Green (PG) Bright Quasar
Survey (BQS) with redshifts in the range $0.04 < z < 0.17$ are presented.
These observations represent the first step towards establishing with a
complete sample whether or not quasi-stellar objects (QSOs) reside in
molecular gas-rich galaxies, as well as towards determining how the
infrared and molecular gas properties of QSOs compare with those of
ultraluminous infrared galaxies (ULIGs), which are a possible evolutionary
precursor of QSOs.  The sample consists of QSOs having absolute blue
magnitudes, $M_{\rm B} \lesssim -22.0$, and infrared excesses, $L_{\rm IR}
(8-1000\micron) /L_{\rm bbb} (0.1-1.0\micron) >0.36$, where the
contribution to the bolometric luminosity of infrared thermal dust
emission for all PG QSOs is typically 20--40\%.  Six out of 10 of the QSOs
observed are detected in the CO($1\to0$) emission line; two detections
confirm previous, less sensitive detections of CO($1\to0$) in PG 1613+658
and PG 0838+770, and four additional QSOs are detected for the first time
(PG 1119+120, PG 1351+640, PG 1415+451, and PG 1440+356).  These six
detections, plus two previous detections of CO in IZw1 and Mrk 1014, bring
the total number of $0.04<z<0.17$ infrared-excess PG QSOs detected in CO
to date to eight, and provide possible evidence that, in addition to
fueling star formation, molecular gas may also serve as a primary source
of fuel for QSO activity.

Both the eight QSOs detected in CO and the four QSOs with non-detections
have high infrared-to-CO luminosity ratios, $L_{\rm IR}/ L'_{\rm CO}$,
relative to most infrared luminous galaxies of the same $L_{\rm IR}$.  The
placement of these QSOs on the $L_{\rm IR} / L'_{\rm CO} - L_{\rm IR}$
plane may be due to significant contributions to dust heated by the QSO in
their host galaxies, due to dust heated by massive stars formed with high
efficiency (i.e., per unit molecular gas mass) relative to most infrared
luminous galaxies, or a combination of both.  If the observed high values
of $L_{\rm IR} / L'_{\rm CO}$ are primarily due to dust heating by QSOs, a
significant fraction of ULIGs with similar values of $L_{\rm IR} / L'_{\rm
CO}$ may also contain buried AGN.  Alternatively, if high $L_{\rm
IR}/L'_{\rm CO}$ is due primarily to star formation, then an enhanced star
formation rate may be intimately connected to the QSO phenomenon.

A comparison of the infrared and CO luminosities of the eight detected and
four undetected QSOs with the optical morphologies of their host galaxies
shows that the three QSOs with $L_{\rm IR}$ and $L'_{\rm CO}$ similar to
ULIGs appear to reside in morphologically disturbed galaxies (i.e.,
ongoing major mergers involving two or more gas-rich disk galaxies),
whereas the host galaxies of the remaining eight QSOs with lower $L_{\rm
IR}$ and $L'_{\rm CO}$ appear to be a mixture of barred spiral host
galaxies, elliptical galaxies, galaxies with an indeterminant
classification, and at least one ongoing major merger.

\end{abstract}

\keywords{quasars: general---galaxies: ISM---infrared: galaxies---ISM:
molecules---radio lines: galaxies---galaxies: active}

\section{Introduction}

There exists substantial evidence that the bulges of nearby massive
galaxies contain quiescent supermassive nuclear black holes ($M_\bullet
= 10^{6-9}$ M$_\odot$: Magorrian et al. 1998; van der Marel 1999;
Kormendy et al. 1998). The almost ubiquitous presence of such objects,
in combination with the possibility that nuclear black hole mass scales
with stellar bulge mass (i.e., the Magorrian Relation: Magorrian et al. 1998),
indicates that supermassive black holes are an essential part of 
spheroid formation, and that mass accretion onto nuclear black holes during
stellar bulge formation likely resulted in periods of AGN activity.

There also exists growing observational evidence that core\footnote{We
define the ``core'' to be the region of a galaxy internal to the radius at
which the projected surface brightness is half its central brightness.
This differs from the ``bulge'' of a galaxy, which is the region internal
to the radius which encloses half the total luminosity of a galaxy.}
formation and AGN activity may be linked with ultraluminous infrared
galaxy (ULIG)\footnote{ULIGs are defined as galaxies with $L_{\rm IR}
[8-1000 \mu{\rm m}] \gtrsim 10^{12}$ L$_\odot$.} mergers, lending support
to the evolutionary connection between ULIGs and QSOs first proposed by
Sanders et al.  (1988a). The evidence can be summarized as follows:
First, observations of massive concentrations of molecular gas in the
nuclear regions of local ULIGs relative to that observed in normal
galaxies (e.g. Kormendy \& Sanders 1992; Downes \& Solomon 1998; Bryant \&
Scoville 1999) support the highly dissipative nature of ULIGs.  Second,
spectroscopic evidence that a significant fraction of ULIGs, most of which
have ``warm'' infrared colors (i.e., 25$\mu$m to 60$\mu$m flux density
ratio, $f_{25}/f_{60} \gtrsim 0.20$), have Seyfert-like emission line
spectra (Veilleux et al. 1995; Veilleux, Kim, \& Sanders 1999) may
indicate that merger events are intimately connected with AGN activity.
Third, recent CO($1\to0$) observations show the association of molecular
gas with the AGN nucleus of several double nucleus ULIGs (Evans et al.
1999; Evans et al. 2000), supporting the necessity of dissipative material
for fueling AGN activity.  Fourth, the recent evidence of similarities in
the space densities of high-redshift submillimeter (SCUBA) galaxies (e.g.,
Smail, Ivison, \& Blain 1997; Barger et al. 1998; Hughes et al. 1998;
Barger, Cowie, \& Sanders 1999) and local giant elliptical galaxies has
been used to infer that ULIGs are connected with massive galaxy formation
(Barger et al. 1999; Trentham 2000).\footnote{This assumes that
high-redshift infrared luminous galaxies are distant analogs of local
infrared galaxies.}

{}From the above, it seems reasonable to assume that a significant fraction
of QSOs may have molecular gas rich hosts and that molecular gas may play
an important role in fueling not only circumnuclear starbursts, but also
AGN activity in galaxies experiencing dissipative collapse.  As a first
step towards establishing the presence of abundant reservoirs of molecular
gas in QSOs, a survey of CO($1\to0$) emission in a complete sample of
low-redshift, optically-selected QSOs has been initiated. In addition to
detecting the possible fueling source for both AGN and star formation
activity, these observations provide a major missing link in comparing the
properties of low-redshifts QSOs and infrared luminous galaxies with both
``cool'' (i.e., 25$\mu$m to 60$\mu$m flux density ratio, $f_{25}/f_{60} <
0.20$) and ``warm'', Seyfert-like ($f_{25}/f_{60} \gtrsim 0.2$) mid-to-far
infrared flux density ratios. Specifically, these data make it possible to
determine whether the relative molecular gas properties of PG QSOs and
ULIGs are consistent with the hypothesis that cool and warm ULIGs
represent successive stages in the evolution toward UV-excess QSOs
(Sanders et al.  1988a). The CO data presented here are complemented by
previous studies of the far-infrared properties (Sanders et al. 1989a) and
host galaxy morphologies (Bahcall et al. 1997; Surace 1998; Surace \&
Sanders 2000:  hereafter SS00) of optical QSOs, and thus facilitate a more
well-rounded comparison with the wealth of data available of low-redshift
($z<0.3$) infrared luminous galaxies.

This paper is divided into 6 sections. Section 2 is a summary of the
sample selection. In \S 3, the general observing procedure is discussed,
followed by the results in \S 4. A comparison of the CO and infrared
properties of the PG QSOs with those of local infrared luminous galaxies
and high redshift QSOs is provided in \S 5, along with a summary of
molecular gas and infrared properties of the PG QSOs relative to the optical
morphologies of their host galaxies.  Finally, \S 6 summarizes the paper.
Throughout this paper, $H_0 = 75$ km s$^{-1}$ Mpc$^{-1}$ and $q_0 = 0.5$
are assumed.

\section{Sample Selection}

The QSOs observed as part of this CO study were selected from a complete
sample of 18 QSOs defined by SS00.  The reader is referred to SS00 for a
detailed discussion of the sample selection, but it is briefly summarized
here for completeness. The selection criteria consisted of all $0.04 < z <
0.17$ QSOs from the Palomar--Green Bright Quasar Survey (PG BGS: Schmidt
\& Green 1983) having infrared-to-``big blue bump'' luminosities, $L_{\rm
IR} (8-1000\micron) / L_{\rm bbb} (0.1-1.0\micron) > 0.36$ and absolute
blue magnitudes, $M_{\rm B} \lesssim -22.0$. The former criteria selects
QSOs with infrared excesses that overlap with warm infrared luminous
galaxies (see \S 5.2).

There are two advantages to using the SS00 sample.  First, the
far-infrared flux densities of these QSOs have been measured using the
Infrared Astronomical Satellite (IRAS) data (Sanders et al. 1989a),
making direct comparisons of their infrared and molecular gas properties
with those of infrared luminous galaxies possible. Second, the molecular
gas properties of these QSOs can be compared with pre-existing optical
images of the QSO host galaxies to search for correlations of $L'_{\rm
CO}$ with the morphology of the host galaxies.

While QSOs with infrared excesses may at first seem to be unrepresentative
of optical QSOs, it must be remembered that at least two-thirds of all the
PG QSOs in the Bright Quasar Survey show clear evidence of infrared
emission consistent with thermal dust emission, with the infrared
components contributing 20-40\% of their total bolometric luminosity
(Sanders et al. 1989a). Thus, the sample selected for this survey is
simply the optical QSOs with the highest probable dust content.

Of the 18 QSOs in the SS00 sample, two of them (IZw1 and Mrk 1014)
have previous confirmed CO($1\to0$) detections (e.g.  Sanders, Scoville,
\& Soifer 1988; Barvainis et al. 1989; Solomon et al.  1997; Schinnerer,
Eckart, \& Tacconi 1998) and two others (PG 0838+770 and PG 1613+658)
have unconfirmed CO detections (Alloin et al. 1992). In addition to
observing PG 0838+770 and PG 1613+658, eight PG QSOs with $L_{\rm IR}
/ L_{\rm bbb}$ ratios  in the range of 0.36--1.1 were selected for CO
observations. The sample is listed in Table 1.

\section{Observations}

Aperture synthesis maps of CO($1\to0$) emission in the sample of PG QSOs
were made with the Owens Valley Radio Observatory (OVRO) Millimeter
Array during 21 observing periods from 1999 April to 2000 October.
The array consists of six 10.4m telescopes, and the longest observed
baseline was 103m. Each telescope was configured with 120$\times$4
MHz digital correlators. All but one of the observations were done
in the low-resolution configuration, providing a $\sim4.0\arcsec$
(full width at half the maximum intensity, or FWHM) synthesized beam
with natural weighting. For each QSO, the receivers were tuned to the
expected frequency of the redshifted CO emission line.  These redshifts
were estimated from published optical emission line redshifts compiled by
the NASA Extragalactic Database (NED); typically, the average or the most
common redshift was adopted.  During the observations, nearby quasars
were observed every 25 minutes to monitor phase and gain variations,
and several brighter quasars were observed to determine the passband
structure.  Observations of Uranus were also made for absolute flux
calibration; typical uncertainties in the flux calibration are 15-20\%.
A journal of observations is provided in Table 2.

\section{Results}

The data were reduced and calibrated using the standard Owens Valley
array program MMA (Scoville et al. 1992). The data were then exported
to the mapping program DIFMAP (Shepherd, Pearson, \& Taylor 1995) and
to the NRAO software package AIPS for spectral line extraction.

Figure 1 shows the integrated CO($1\to0$) intensity maps for the detected
PG QSOs (integrated intensity values are listed in Table 3).  Six out of
the ten QSOs observed were detected in CO; all of the detections have a
signal-to-noise ratio, S/N, greater than 4.5$\sigma$.  A comparison was
done between the CO beam centroid positions and the optical QSO positions,
the latter of which were extracted from the USNO-A2.0 Catalog. With
the exception of PG 1351+640 (1.44\arcsec$~$ offset) and PG 1613+658
(1.05\arcsec$~$ offset), all the CO centroid and optical positions agreed
to within 0.9\arcsec. Of the four non-detections, two are of the QSOs
with the highest redshifts in the sample (PG 1402+261 and PG 1202+281).
For all four  non-detections, the 1.0 GHz continuum channel data was
examined to search for CO emission over a broader (albeit less sensitive)
bandwidth than covered by the spectrometer. Such searches yielded null
results.

In the present sample of ten QSOs, PG 0007+106 is the only one having
detectable 3mm continuum emission.  The unresolved continuum emission is
observed to be at RA = 00:10:30.98 dec = +10:58:29.8 (J2000.0), which is
within 0.75$\arcsec$ of the USNO-A2.0 optical position, and the continuum
flux density is measured to be 0.1 Jy.

Figure 2 shows the CO($1\to0$) spectra of the six detected QSOs.  The CO
emission line widths are varied, ranging from FWHM of $\Delta v_{\rm FWHM}
\sim 50-90$ km s$^{-1}$ for PG 1415+451 and PG 0838+770 to $\sim 500$ km
s$^{-1}$ for PG 1613+658. All of CO emission line centroids are within 300
km s$^{-1}$ of the adopted optical redshifts, indicating that the optical
emission lines are only moderately red/blueshifted relative to the
systemic velocity.  Both PG 0838+770 and PG 1613+658 are detected in CO;
the emission line widths and redshifts are similar to those reported by
Alloin et al.  (1992), however, the integrated CO line intensity of PG
0838+770 is only 40\% of that reported by Alloin et al. (1992). By
comparison, the CO line intensity of PG 1613+658 agrees with the Alloin et
al. (1992) measurement to within 12\% (i.e.  within the error of the OVRO
measurements).

While the CO emission line of PG 1415+451 was detected with a
signal-to-noise of 5.2$\sigma$, the narrow appearance of the emission
feature may be an indication that the ``line'' is actually an artifact;
i.e., a spurious interference spike created at the phase center of the
interferometer. We believe that the emission feature is authentic based on
a combination of the following three facts: First, the emission feature
spans five independent (i.e., unbinned) channels. Second, 
while the phase center of the PG 1415+451 observations is within the 
beam of the emission feature, and the centroid
position of the feature is $\sim0.9\arcsec$ closer to the optical QSO
position (obtained from the USNO--A2.0 catalog) than the phase center is
to the optical QSO position.  Third, the CO observation of PG 0838+770,
which has a CO line width less than a factor of two wider than the
emission feature observed for PG 1415+451, was a confirmation of prior
observations reported by Alloin et al. (1992), and is thus authentic.  It
is thus likely that the host galaxies of both QSOs simply have molecular
disks that are being observed nearly face-on.

\subsection{CO Line Luminosities and Molecular Gas
Masses}

Table 3 summarizes the emission line properties of the ten PG QSOs.
The luminosity distance for a source at a given redshift is,
$$D_{\rm L} = cH^{-1}_0q^{-2}_0 \left\{ z q_0 + (q_0 - 1)
\left( \sqrt{2 q_0 z + 1} - 1 \right) \right\}$$
$$[{\rm Mpc}].
\eqno(1)$$
Given the measured flux,
$S_{\rm CO} \Delta v$,
the CO luminosity of a source at
redshift $z$ is,
$$L'_{\rm CO} = \left( {c^2 \over {2 k \nu^2_{\rm obs}}} \right)
S_{\rm CO} \Delta v D^2_{\rm L} (1 + z)^{-3}, \eqno(2)$$
where $c$ is the speed of light, $k$ is the Boltzmann constant,
and $\nu_{\rm obs}$ is the observed frequency.
In terms of useful units,
$L'_{\rm CO}$
can be written as,
$$L'_{\rm CO} = 1.35\times10^3
\left( S_{\rm CO} \Delta v \over {\rm Jy~km~s}^{-1} \right)
\left( D^2_{\rm L} \over {\rm Mpc^2} \right) J^{-2} (1 + z)^{-1}$$
$$[{\rm K~km~s}^{-1} {\rm~pc}^2]. \eqno(3)$$

To calculate the mass of molecular gas in the PG QSOs, the reasonable
assumption is made that the CO emission is optically thick and thermalized
(see Discussion Section), and that it originates in gravitationally
bound molecular clouds. Thus, the ratio of the H$_2$ mass and the CO
luminosity is given by,

$$\alpha = {M({\rm H}_2) \over L^\prime_{\rm CO}} \propto {\sqrt {n({\rm H}_2)}
\over 
T_{\rm b}} {\rm M}_\odot ~~[{\rm ~K~km~s}^{-1} {\rm ~pc}^2]^{-1}, \eqno(4)$$

\noindent
where
$n($H$_2)$ and $T_{\rm b}$ are the density of H$_2$ and brightness
temperature for the CO(1$\to$0) transition (Scoville \& Sanders 1987;
Solomon, Downes, \& Radford 1992).  Multitransition CO surveys of
molecular clouds in the Milky Way (e.g. Sanders et al. 1993), and in
nearby starburst galaxies (e.g. G\"{u}sten et al. 1993) have shown that
hotter clouds tend of be denser such that the density and temperature
dependencies tend to cancel each other. The variation in the value of
$\alpha$ is approximately a factor of 2 for a wide range of kinetic
temperatures, gas densities, and CO abundance (e.g. $\alpha = 2-5
M_{\odot}$ [K km s$^{-1}$ pc$^2]^{-1}$:  Radford, Solomon, \& Downes
1991).  More recent dynamical mass estimates of a low-redshift infrared
galaxy sample observed in CO with the Plateau de Bure Interferometer
indicate that $\alpha$ can be as low as 1 $M_{\odot}$ [K km s$^{-1}$
pc$^2]^{-1}$ (Downes \& Solomon 1998).  We adopt a value of 4 $M_{\odot}$
(K km s$^{-1}$ pc$^2)^{-1}$ for $\alpha$, which is similar to the value
determined for the bulk of the molecular gas in the disk of the Milky Way
(Scoville \& Sanders 1987; Strong et al. 1988).

In the four cases where only upper limits to the CO flux were determined,
the average FWHM velocity of all the PG QSOs detected in CO to date was
adopted (280 km s$^{-1}$). Conservative upper limits to the CO luminosity
and molecular gas mass were then calculated from the 3$\sigma$ upper
limits to the CO flux density.

\section{Discussion}

The detection of CO emission in the host galaxies of these PG QSOs
provides direct evidence that these infrared-excess QSOs live in
host galaxies containing significant supplies of molecular gas. It
also supports the hypothesis that galaxies with a molecular gas-rich
interstellar medium (ISM) may be capable of igniting and sustaining both
AGN and star formation activity in QSO host galaxies. The measured CO
luminosities for the six detected QSOs in Table 3 translate into molecular
gas masses of $M({\rm H}_2) \sim 2-20\times10^9$ M$_\odot$, which, along
with the additional two QSOs IZw1 and Mrk 1014, span the full range
of masses derived for luminous infrared galaxies (LIGs)\footnote{LIGs
are defined as galaxies having $L_{\rm IR} [8-1000 \mu{\rm m}] =
10^{11.0-11.99}$ L$_\odot$.} and ULIGs.

These six PG QSOs are among the faintest CO($1\to0$) sources yet detected
at redshifts $z < 0.3$.  An interesting issue is whether or not the CO
luminosities of these sources, and perhaps the CO luminosities of other PG
QSOs as well, could have been predicted apriori. Consider the plot of CO
flux versus the 100$\mu$m flux density of the PG QSO sample, in addition
to a comparative sample of $z < 0.3$ warm and cool ULIGs observed in CO to
date (Figure 3). Solomon et al.  (1997) use the tightness of this
correlation, in combination with the small sizes of the CO emitting
regions in ULIGs, to argue that the dust emission at 100$\mu$m and the
CO(1$\to$0) emission is optically thick (see their Figure 4). The $4\arcsec$
resolution of the present CO dataset is insufficient to determine
how compact the CO emitting regions in these PG QSOs are (e.g., $4\arcsec$
corresponds to $\sim 6$ kpc at $z \sim 0.1$). For our purposes, the fact
that all of the PG QSOs detected in CO to date also fall along this
correlation indicates that the CO fluxes (and thus, the molecular gas
masses) of optical QSOs may potentially be estimated from the 100$\mu$m
flux density, $f_{100}$, to within a factor of a few if $f_{100}$ is known
to high accuracy.  Note, however, that the error in $f_{100}$ of any {\it
IRAS} measurements exceeds 25\% below 0.5 Jy; improved far-infrared
measurements of QSOs will be possible via observations with the Space
Infrared Telescope Facility (SIRTF).

With the detection of molecular gas in the six of the ten optical QSOs
listed in Table 3, plus the previous detection of IZw1 and Mrk 1014,
physically meaningful comparisons between QSOs and infrared-selected
galaxies can be better addressed.  The remainder of the discussion section
will thus be focussed on comparing the infrared and CO luminosities of
the eight QSOs detected in CO to date with infrared luminous galaxies,
followed by a brief discussion of recent images of the host galaxies
of these PG QSOs (Bahcall et al.  1997; SS00).  The four QSOs with CO
non-detections will be discussed where applicable.

\subsection{Comparison with Infrared Luminous Galaxies}

Figure 4 is a plot of CO luminosity versus redshift of the 12 QSOs.  In
order to examine how the CO luminosity of these PG QSOs compares with that
of the most CO luminous galaxies presently known at any given redshift,
the infrared luminous galaxies observed in CO($1\to0$) by the Sanders et
al. (1989b), Sanders, Scoville, \& Soifer (1991), and Solomon et al.
(1997) surveys have also been plotted. All but eight of the infrared
luminous galaxies at $z > 0.04$ are ULIGs.  Three features of Figure 4 are
readily apparent.  First, the CO luminosity range of all twelve QSOs spans
more than an order of magnitude. This range is significantly more than
that of the cool ULIGs, all of which have $L'_{\rm CO} \sim 10^{10}$ K km
s$^{-1}$ pc$^2$ over the same redshift range. Second, the three QSOs with
the highest $L'_{\rm CO}$ (IZw1, PG 1613+658, and Mrk 1014) fall within
the scatter of the cool ULIGs distribution, whereas the remaining nine
QSOs form a distinct group with lower values of CO luminosity for their
given redshift.  Third, seven of the eight QSOs detected in CO have a
range in $L'_{\rm CO}$ similar to that of the warm ULIGs. 
Thus, while the $L'_{\rm CO}$ values of the QSOs are diverse,
there is also evidence of overlap in $L'_{\rm CO}$ between these
optically-selected and infrared-selected samples.

Consider now the infrared luminosity of both samples relative to
the CO luminosity.  Figure 5 shows two ways in which infrared and CO
data of infrared luminous galaxies are commonly plotted. The top panel
(Figure 5a) is a plot of $L_{\rm IR}$ versus $L'_{\rm CO}$, which shows
the relationship between the fuel available for star formation and AGN
activity ($L'_{\rm CO}$) and the luminosity of the imbedded sources
($L_{\rm IR}$). The bottom panel (Figure 5b) is a plot of $L_{\rm
IR}/L'_{\rm CO}$ versus $L_{\rm IR}$. In starburst galaxies, $L_{\rm
IR}/L'_{\rm CO}$ is commonly referred to as the star formation efficiency,
it is a measure of the number of high-mass stars (as traced by $L_{\rm
IR}$) created per unit molecular gas mass (as traced by $L'_{\rm CO}$).
An examination of Figure 5a shows that the quantity $L_{\rm IR}$ increases
as $L'_{\rm CO}$ increases, and for any given CO luminosity bin, the QSOs
are in the upper half of the distribution. In Figure 5b, the QSOs populate
the upper half of the LIG-ULIG distribution of $L_{\rm IR}/L'_{\rm CO}$.
Note also that, in general, the warm infrared galaxies and QSOs populate
similar areas of both plots.

The interpretation of Figure 5 depends on whether the infrared emission
in QSOs is due to dust heating by AGN or massive stars.  First, consider
the former possibility: Sanders et al. (1988a) proposed that ULIGs are
dust-enshrouded QSOs; in this scenario, the optical QSO is revealed after
most of the molecular gas is consumed by star formation and AGN activity,
and after most of the dust is destroyed/blown out by winds associated with
supernovae and AGN.  During this process, ULIGs pass from a cool infrared
phase where vigorous star formation and black hole creation/accretion
has commenced, through a warm infrared phase (see also Sanders 1988b).
This idea is further addressed by Sanders et al. (1989a), who propose
that the infrared emission from PG QSOs is thermal emission from dust
that has absorbed 20--40\% of the emission emanating from the QSO
(i.e., $L_{\rm IR} = [0.2-0.4] L_{\rm bol}$, where $L_{\rm bol}$ is the
bolometric luminosity).  Given this, the quantity $L_{\rm IR}/L'_{\rm
CO}$ is not a measure of the star formation efficiency in QSOs, but is
an indication that the AGN is contributing significantly to heating the
dust in QSOs (see also Evans et al. 1998).

If the dust-enshrouded quasar model of ULIGs is correct and the ratio
$L_{\rm IR}/L_{\rm bol}$ for both ULIGs and QSOs is due to covering factor
and not due to orientation effects (e.g. Barthel 1989), optically-selected
QSOs must have had a covering factor of unity in their past ULIG phase.
Consider now where the QSOs would be on Figure 5a and 5b if the QSO nuclei
were completely enshrouded in dust (i.e.  $L_{\rm bol} = L_{\rm IR}$).
This increases $\log L_{\rm IR}$ of the QSOs by $\sim$0.3--0.5 in Figure
5a, and shows to first order a possible range in locations of enshrouded
QSOs on this diagram.  In Figure 5b, the quantity $L_{\rm IR}/L'_{\rm
CO}$ of the QSOs is increased to extreme values ($\sim 400-1600$). Note
that, in both panels, ULIGs with high ($>160$) $L_{\rm IR}/L'_{\rm CO}$
occupy the region covered by the QSOs, similar to what would be expected
if these ULIGs were dust-enshrouded QSOs.

As mentioned above, the second possible interpretation of the infrared
emission in QSOs is that $L_{\rm IR}$ is primarily due to dust heating by
massive stars.  Thus, high values of $L_{\rm IR}$ (Figure 5a) and $L_{\rm
IR}/L'_{\rm CO}$ (Figure 5b) would result from an extremely efficient
production of high-mass stars per unit molecular gas mass in QSOs relative
to most LIGs and ULIGs. The attraction of this scenario is that both stars
forming in the central regions of galaxies and AGN contribute
significantly to the overall energy budget of their host galaxies, and
that AGN activity is somehow strongly dependent on an accompanying
starburst.  Indeed, observed correlations between the masses of quiescent
supermassive nuclear black holes and both the stellar bulge masses (e.g.,
Magorrian et al. 1998) and velocity dispersions (Ferrarese \& Merritt
2000; Gebhardt et al. 2000) of nearby normal galaxies suggests that the
accretion of mass onto nuclear black holes may accompany stellar bulge
formation.  However, whether or not a significant fraction of the stellar
bulge population of the present sample of PG QSOs is in the process of
being formed cannot be addressed with the present data set.

\subsection{The Host Galaxies}

An obvious issue to consider at this point is whether or not the diversity
of $L'_{\rm CO}$ and $L_{\rm IR}$ of these 12 QSOs is at all correlated
with the galaxy type or evolutionary state of their host galaxies.
An examination of optical and near-infrared images of all 12 QSOs
(Bahcall et al. 1997; Surace et al.  1998; SS00) reveals that two of
the QSOs with the highest $L'_{\rm CO}$ (Mrk 1014 and PG 1613+658) have
irregular morphologies consistent with an advanced merger of two molecular
gas-rich disk galaxies (see simulations by Barnes \& Hernquist 1996), and
the other (IZw1) has asymmetric spiral arms, suggestive of a strong
interaction with a much smaller mass companion. The remaining nine QSOs
have host galaxies that appear to be a mixture of barred spiral
galaxies, elliptical galaxies, galaxies with indeterminant features, and
at least one ongoing merger (i.e., PG 0007+106).

The morphologies are suggestive of a diversity in paths that may
ultimately lead to QSO activity (SS00).  First, the QSOs having the
highest values of $L'_{\rm CO}$ appear to be ongoing major mergers
(collisions of two or more molecular gas-rich disk galaxies where
prominent tidal features are still visible). Thus, in terms of $L'_{\rm
CO}$, $L_{\rm IR}$, and host morphologies, these QSOs are similar to
ULIGs.  Second, the less CO-luminous QSOs with barred spiral galaxy hosts
have a less clear origin, however, the bar likely acts as the means by
which molecular gas in the spiral arms is torqued and drained of enough
angular momentum to agglomerate in the nuclear regions (e.g. Sakamoto et
al.  1999).  These barred spiral galaxies may have lower $L'_{\rm CO}$
because only one, massive molecular gas-rich galaxy is involved (though
the bar instability may be triggered via an interaction with a satellite
galaxy).  Third, the QSOs having host galaxies with an indeterminant
classification may be spiral galaxies with features too faint to discern
with the sensitivity of the SS00 observations, fully evolved elliptical
galaxies, or well-evolved major mergers which are evolving into elliptical
galaxies (see simulations by Barnes \& Hernquist 1996).  Thus, these host
galaxies may have low $L'_{\rm CO}$ because the host galaxy is
intrinsically molecular gas poor, or because the molecular gas has already
been consumed via both circumnuclear star formation and the QSO.
Finally, the QSOs with elliptical host galaxies may be the end stage of the
merger one two disk galaxies, or elliptical host galaxies that have
cannibalized a satellite galaxy (such as the $z\sim0.44$ 
galaxy IRAS PSC 09104+4109: e.g. Evans et al. 1998).

\subsection{Comparison with High Redshift QSOs}

A significant number of QSOs have been detected at high redshift ($z>2$:
Barvainis et al. 1994, 1998; Ohta et al. 1996; Omont et al. 1996;
Guilloteau et al. 1997; Downes et al. 1999).  These QSOs, many of which
were initially selected for CO observations based on their strong
rest-frame infrared luminosity, are likely among the most molecular gas
rich QSOs at their redshift.  Figure 6 is a plot of the CO luminosity of
the eight PG QSOs detected in CO to date, as well as all of the moderate
($0.3<z\lesssim2$) and high-redshift CO luminous QSOs, versus the
redshift.  Several of the QSOs at high redshift are known to be
gravitationally lensed; in these cases, both observed and the intrinsic CO
luminosities are plotted.  Based on the present sample, there is no
convincing evidence for evolution of molecular gas mass of the most CO
luminous QSOs at any given redshift as a function of increasing redshift.
The ratio of $L'_{\rm CO}$ of BRI 1335-0415 ($z=4.7$) to that of the most
CO luminous low-redshift QSO (Mrk 1014: $z=0.16$) is three.  However,
given the uncertainty in the $L'_{\rm CO}$--to--$M($H$_2)$ conversion
factor, $\alpha$, the factor of three is also consistent with no
evolution.

The derived molecular gas masses of the most CO luminous QSOs at any given
redshift are significantly below the average stellar mass of present-day
elliptical galaxies ($\sim$ several$\times10^{11}$ M$_\odot$). Given that
massive elliptical galaxies are a likely end stage for many QSO host
galaxies (e.g., as inferred from Bahcall et al. 1997 or the Magorrian
Relation), the mass discrepancy can be rectified by concluding that
present-day massive galaxies are created from the merger of smaller
galaxies, and/or that galaxies with molecular gas masses in excess of
several$\times10^{10}$ M$_\odot$ quickly consume the gas by forming high
mass stars (e.g. Evans et al. 1996, 1998).

\section{Summary}

Millimeter (CO) observations of 10 infrared-excess, optically-selected
Palomar-Green QSOs with redshifts in the range $0.04 < z < 0.17$ have been
presented.  These observations represent the first step towards detecting
molecular gas in a complete sample of optical QSOs. The following
conclusions have been reached:

1) Six out of 10 QSOs in the sample were detected in CO$(1\to0)$.
The CO detections in PG 1613+658 and PG 0838+770 confirm previously
reported detections by Alloin et al. (1992). These detections, combined
with previous detections of CO in the PG QSOs IZw1 and Mrk 1014, provide
strong support for the possibility that molecular gas is a likely source
of fuel for both AGN and star formation in QSO host galaxies.

2) All of the QSOs have high values of $L_{\rm IR}/ L'_{\rm CO}$ relative
to most $z<0.3$ infrared luminous galaxies of the same $L_{\rm IR}$.
This is because QSOs contribute significantly to the heating of dust
in their host galaxies, and/or because the dust is heated by massive
stars formed with high efficiency (i.e., per unit molecular gas mass)
relative to most infrared luminous galaxies.  If large values of $L_{\rm
IR} / L'_{\rm CO}$ are achieved primarily through dust heating by QSOs,
a significant fraction of ULIGs with similar values of $L_{\rm IR} /
L'_{\rm CO}$ may also contain buried AGN.  Alternatively, if high $L_{\rm
IR}/L'_{\rm CO}$ is due primarily to star formation, then high-mass star
formation may be intimately connected to the QSO phenomenon.

3) The warm infrared luminous galaxies, which by definition have
Seyfert-like mid-to-far infrared flux density ratios (i.e., $f_{25}/f_{60}
\gtrsim 0.2$), have similar ranges of $L_{\rm IR}/ L'_{\rm CO}$ and
$L'_{\rm CO}$ as the infrared-excess PG QSOs.

4) A comparison of the CO luminosities and optical morphologies of the
QSO host galaxies indicates that QSOs with the highest infrared and CO
luminosities reside in morphologically disturbed galaxies, whereas the
rest have a mixture of barred spiral host galaxies and ``featureless''
host galaxies.

5) There is no convincing evidence of evolution in the molecular gas
content of the most CO luminous QSOs in the redshift range $z=0-5$.
The most CO luminous high-redshift QSO (BRI 1335-0415) has an $L'_{\rm
CO}$ a factor of three greater than the most CO luminous low-redshift
QSO (Mrk 1014).  However, given the uncertainty in the $L'_{\rm
CO}$--to--$M($H$_2)$ conversion factor, $\alpha$, the factor of three
is also consistent with no evolution.

\acknowledgements

We thank the staff and postdoctoral scholars of the Owens Valley
Millimeter array for their support both during and after the observations
were obtained. ASE also thanks J. Carpenter and A. Baker for useful
discussions and assistance during the preparation of this manuscript, and
the referee for comments that led to the clarification of several key
points.  ASE was supported by RF9736D and AST 0080881. JAS was supported
by the Jet Propulsion Laboratory, California Institute of Technology,
under contract with NASA. The Owens Valley Millimeter Array is a radio
telescope facility operated by the California Institute of Technology and
is supported by NSF grant AST 9981546.  This research has made use of the
NASA/IPAC Extragalactic Database (NED) which is operated by the Jet
Propulsion Laboratory.



\vskip 0.3in
\centerline{Figure Captions}

\vskip 0.2in

\noindent
Figure 1. Integrated intensity maps of the six PG QSOs detected
in CO($1\to0$). a) PG 0838+770 - contours are plotted as 1$\sigma
\times (-1.6, 1.6, 2.6, 3.6, 4.6, 5.6, 6.6)$. The peak intensity is
0.017 Jy beam$^{-1}$ and corresponds to the position RA=08:44:45.22
dec=+76:53:08.96 (J2000.0).
b) PG 1119+120 - contours are plotted as 1$\sigma \times
(-2.3, 2.3, 3.3, 4.3, 5.3)$. The peak intensity is 0.014 Jy beam$^{-1}$,
and corresponds to the position RA=11:21:47.12 dec=+11:44:18.30 (J2000.0).
c) PG 1351+640 - contours are plotted as 1$\sigma \times
(-1.6, 1.6, 2.6, 3.6, 4.6)$; the peak intensity is 0.0055 Jy beam$^{-1}$,
and corresponds to the position RA=13:53:15.62 dec=+63:45:45.72 (J2000.0).
d) PG 1415+451 - contours are plotted as 1$\sigma \times
(-2.2, 2.2, 3.2, 4.2, 5.2)$; the peak intensity is 0.051 Jy beam$^{-1}$,
and corresponds to the position RA=14:17:00.76 dec=+44:56:06.50 (J2000.0).
e) PG 1440+356 - contours are plotted as 1$\sigma \times
(-2.3, 2.3, 3.3, 4.3, 5.3, 6.3)$; the peak intensity is 0.016 Jy beam$^{-1}$,
and corresponds to the position RA=14:42:07.48 dec=+35:26:22.33 (J2000.0).
f) PG 1613+658 - contours are plotted as 1$\sigma \times
(-1.6, 1.6, 2.6, 3.6, 4.6, 5.6, 6.6, 7.6, 8.6)$; the peak intensity
is 0.012 Jy beam$^{-1}$,
and corresponds to the position RA=16:13:57.15 dec=+65:43:09.62 (J2000.0).

\vskip 0.1in
Figure 2. CO($1\to0$) spectra of the six detected PG QSOs. For PG 0838+770
and PG 1415+451, the data have been smoothed to a resolution of 16 MHz
with a 8 MHz sampling. For PG 1119+120, the data have been smoothed to
a resolution of 32 MHz with a 16 MHz sampling. For PG 1351+640, the data
have been smoothed to a resolution of 40 MHz with a 20 MHz sampling. For
PG 1440+356 and PG 1613+658, the data have been smoothed to a resolution
of 24 MHz with a 12 MHz sampling.

\vskip 0.1in
Figure 3. A plot of the CO flux vs. the 100 $\mu$m flux density for PG
QSOs and ULIGs detected to date in CO. The CO data for IZw1 and Mrk 1014
were obtained from Barvainis et al. (1989) and Solomon et al.  (1997),
respectively. The ULIGs CO data were obtained from Sanders et al. (1989b),
Sanders, Scoville, \& Soifer (1988, 1991), Solomon et al. (1997), and
Evans et al. (1999). The 100$\mu$m flux densities were obtained from
Sanders et al. (1989a, 1991) and Solomon et al. (1997). Arrows denote
3$\sigma$ upper limits on the CO luminosity of PG 1126-041, PG 1202+281,
and PG 1402+261. Only upper limits on $L'_{\rm CO}$ and $f_{100}$
currently exist for PG 0007+106, and thus PG 0007+106 is not included on
the plot. Adapted from Figure 4 of Solomon et al. (1997).

\vskip 0.1in 
Figure 4. A plot of $\log L'_{\rm CO}$ versus redshift ($z$) for the
low-$z$ PG QSO sample, a flux-limited sample ($f_{60} > 5.24$ Jy) of
infrared luminous galaxies and a sample of ultraluminous infrared
galaxies. The vertical dashed lines represent the upper and lower $z$
boundaries of the PG QSO sample. The LIG and ULIGs data have been obtained
from the same sources as in Figure 3.  Arrows denote 3$\sigma$ upper
limits on the CO luminosity of PG 1126-041, PG 1202+281, and PG 1402+261.

\vskip 0.1in
Figure 5. a) A plot of $L_{\rm IR}$ vs. $L'_{\rm CO}$ for the low-$z$ QSO
sample, a flux-limited sample ($f_{60 \mu{\rm m}} > 5.24$ Jy) of infrared
luminous galaxies and a sample of ultraluminous infrared galaxies.  Arrows
denote 3$\sigma$ upper limits on the CO luminosity of PG 1126-041, PG
1202+281, and PG 1402+261.  b) A plot of $L_{\rm IR}/L'_{\rm CO}$ vs.
$L_{\rm IR}$ for the low-$z$ QSO sample, a flux-limited sample ($f_{60
\mu{\rm m}} > 5.24$ Jy) of infrared luminous galaxies and a sample of
ultraluminous infrared galaxies.  Arrows denote 3$\sigma$ lower limits on
$L_{\rm IR}/L'_{\rm CO}$ of PG 1126-041, PG 1202+281, and PG 1402+261.
For simplicity, all of the cool infrared luminous galaxies are plotted
as plus signs.

\vskip 0.1in
Figure 6. A plot of $\log L'_{\rm CO}$ versus redshift ($z$) for the
low-$z$ PG QSO sample and the high-redshift QSOs detected in CO to date.
The moderate and high-redshift QSO data have been obtained from the following
references: 3C 48 ($z = 0.37$: Scoville et al. 1993), H1413+117 ($z =
2.56$: Barvainis et al. 1994), MG 0414+0534 ($z=2.64$: Barvainis et al.
1998), APM 08279+5255 ($z=3.91$: Downes et al. 1999), BRI 1335-0415
($z=4.41$:  Guilloteau et al. 1997), and BR 1202-0725 ($z=4.69$: Omont et
al. 1996).  For gravitationally lensed QSOs, $L'_{\rm CO}$ is plotted in
terms of both the observed value and the intrinsic value. For MG
0414+0534, the amplification is unknown, thus MG 0414+0534 is plotted with
a solid line extending downward from the observed $L'_{\rm CO}$.  Adapted
from Figure 3 of Frayer et al. (1999).

\begin{deluxetable}{lcclccc}
\tablenum{1}
\tablewidth{0pt}
\tablecaption{Source List}
\tablehead{
\multicolumn{1}{c}{} &
\multicolumn{2}{c}{Optical Coordinates$^a$ (J2000.0)} &
\multicolumn{1}{c}{} &
\multicolumn{1}{c}{$\log L_{\rm IR}^{~~~c}$} &
\multicolumn{1}{c}{$\log L_{\rm bol}^{~~~c}$} &
\multicolumn{1}{c}{}\nl
\multicolumn{1}{c}{Source} &
\multicolumn{1}{c}{R.A.} &
\multicolumn{1}{c}{Decl.} &
\multicolumn{1}{c}{$z^{~~~b}_{\rm opt}$} &
\multicolumn{1}{c}{($\log L_\odot$)} &
\multicolumn{1}{c}{($\log L_\odot$)} &
\multicolumn{1}{c}{$L_{\rm IR}/L_{\rm bbb}$}
}
\startdata
PG 0007+106 = IIIZw2 & 00:10:30.98 & +10:58:29.8 & 0.089 & 11.63 & 12.21 & 0.36 \nl
PG 0838+770 & 08:44:45.36 & +76:53:09.2 & 0.131 & 11.60 & 12.01 & 0.63 \nl
PG 1119+120 = Mrk 734$^d$ & 11:21:47.15 & +11:44:19.0 & 0.050 & 11.17 & 11.47 & 1.00 \nl
PG 1126-041 = Mrk 1298 & 11:29:16.72 & -04:24:07.6 & 0.060 & 11.47 & 11.94 & 0.50 \nl
PG 1202+281 & 12:04:42.12 & +27:54:11.8 & 0.165 & 11.78 & 12.07 & 1.05 \nl
PG 1351+640 & 13:53:15.83 & +63:45:45.6 & 0.088 & 11.82 & 12.37 & 0.40 \nl
PG 1402+261 = Ton182 & 14:05:16.21 & +25:55:34.1 & 0.164 & 11.85 & 12.33 & 0.49 \nl
PG 1415+451 & 14:17:00.84 & +44:56:06.5 & 0.114 & 11.48 & 11.83 & 0.80 \nl
PG 1440+356 = Mrk 478 & 14:42:07.48 & +35:26:23.1 & 0.079 & 11.62 & 11.95 & 0.87 \nl
PG 1613+658 = Mrk 876 & 16:13:57.21 & +65:43:10.6 & 0.129 & 11.99 & 12.20 & 1.58 \nl
\enddata
\tablecomments{Two additional PG QSOs were previously detected as part
of a survey of CO emission in ``warm'' ULIGs. They are PG 0157+001 (=
Mrk 1014), which has $L_{\rm IR} = 10^{12.49}$ L$_\odot$, $L_{\rm bol}
= 10^{12.68}$ L$_\odot$, and $L_{\rm IR}/L_{\rm bbb} = 2.57$, and PG
0050+124 (= IZw1), which has $L_{\rm IR} = 10^{12.00}$ L$_\odot$, $L_{\rm
bol} = 10^{12.32}$ L$_\odot$, and $L_{\rm IR}/L_{\rm bbb} = 2.01$.}
\tablenotetext{a}{Optical Coordinates have been extracted from the
USNO-A2.0 Catalog.}
\tablenotetext{b}{Redshifts Based on Optical Emission Lines.}
\tablenotetext{c}{Calculated assuming $H_0 = 75$ km s$^{-1}$ Mpc$^{-1}$
and $q_0 = 0.5$.}
\tablenotetext{d}{While PG 1119+120 has an $L_{\rm IR}/L_{\rm bbb}$ within
the sample selection range, it has an absolute $B$ magnitude of $-21.4$,
which is outside of the range of $M_{\rm B} \lesssim -22.0$ occupied by
the rest of the sample. However, it has historically been designated a QSO
(see SS00) and thus is included in our discussion.}
\end{deluxetable}

\begin{deluxetable}{lllllcl}
\pagestyle{empty}
\scriptsize
\tablenum{2}
\tablewidth{0pt}
\tablecaption{Journal of Observations}
\tablehead{
\multicolumn{1}{c}{} &
\multicolumn{1}{c}{} &
\multicolumn{4}{c}{Phase/Gain Calibrator} &
\multicolumn{1}{c}{} \nl
\cline{3-6} \\
\multicolumn{1}{c}{Source} &
\multicolumn{1}{c}{Dates} &
\multicolumn{1}{c}{} &
\multicolumn{2}{c}{J2000.0} &
\multicolumn{1}{c}{Flux} &
\multicolumn{1}{l}{Passband} \nl
\multicolumn{1}{c}{} &
\multicolumn{1}{c}{} &
\multicolumn{1}{c}{Calibrator} &
\multicolumn{1}{c}{R.A.} &
\multicolumn{1}{c}{Decl.} &
\multicolumn{1}{c}{(Jy)} &
\multicolumn{1}{l}{Calibrator} \nl}
\startdata
PG 0007+106 & 2000 Oct 11   & 0106+013 & 01:08:38.72 & +01:35:00.38 & 1.82 &
3c454.3,3c84\nl
            & 2000 Oct 12   &          & & &     & \nl 
PG 0838+770 & 2000 April 24 & 1044+719 & 10:48:27.63 & +71:43:35.84 & 1.36 & 3C 273 \nl
            & 2000 April 25 &          & & &      & \nl      
PG 1119+120 & 2000 March 24 & 1055+018 & 10:58:29.57 & +01:33:58.83 & 2.11 & 3C 273\nl 
            & 2000 April 19 &          & & &      & \nl
PG 1126-041 & 1999 May 30   & 1055+018 & 10:58:29.57 & +01:33:58.83 & 2.14 & 3C 273 \nl 
            & 1999 May 31   &          & & &      & \nl
PG 1202+281 & 1999 May 02   & 1156+295 & 11:59:31.80 & +29:14:43.37 & 3.41 & 3C 273 \nl 
            & 1999 May 03   &          & & &      & 3C 84,3C 273,3C 454.3 \nl
PG 1351+640 & 1999 May 15   & 1044+719 & 10:48:27.63 & +71:43:35.84 & 0.62 &3C 273 \nl 
            & 1999 May 18   &          & & &      & \nl
            & 1999 May 20   &          & & &      & \nl
PG 1402+261 & 2000 May 21   & 1308+326 & 13:10:28.62 & +32:20:43.42 & 0.74 &3C 273 \nl 
            & 2000 Oct 11   &          & & & 0.68 & 3c273,3c345 \nl
            & 2000 Oct 18   &          & & & 0.68 & 3c273 \nl
PG 1415+451 & 2000 May 07   & 1308+326 & 13:10:28.62 & +32:20:43.42 & 0.67 &3C 273 \nl 
PG 1440+356 & 1999 May 22   & 1611+343 & 16:13:40.99 & +34:12:47.67 & 1.80 &3C 273,3C
454.3 \nl 
            & 1999 June 03  &          & & &      &3C 84,3C 111,3C 273, 3C 345 \nl
PG 1613+658 & 1999 April 19 & 1803+784 & 18:00:45.48 & +78:28:03.85 & 1.69 &3C 273 \nl 
            & 1999 April 25 &          & & &      & \nl
\enddata
\end{deluxetable}

\begin{deluxetable}{lcllcrr}
\tablenum{3}
\tablewidth{0pt}
\tablecaption{CO Emission Line Properties}
\tablehead{
\multicolumn{1}{c}{Source} &
\multicolumn{1}{c}{$D_{\rm L}^{~~~a}$} &
\multicolumn{1}{c}{$z_{\rm CO}$} &
\multicolumn{1}{c}{$\Delta v_{\rm FWHM}$} &
\multicolumn{1}{c}{$S_{\rm CO} \Delta v$} &
\multicolumn{1}{c}{${L'_{\rm CO}}^{c}$} &
\multicolumn{1}{c}{$M$(H$_2)^d$}\nl
\multicolumn{1}{c}{} &
\multicolumn{1}{c}{(Mpc)} &
\multicolumn{1}{c}{} &
\multicolumn{1}{c}{(km s$^{-1}$)} &
\multicolumn{1}{c}{(Jy km s$^{-1}$)} &
\multicolumn{1}{c}{(K km s$^{-2}$ pc$^2$)} &
\multicolumn{1}{c}{($M_\odot$)}
}
\startdata
PG 0007+106 & 360 & \nodata & \nodata & $<3.0^b$ & $<8.8\times10^8$ & $<3.5\times10^9$
\nl
PG 0838+770 & 540 & 0.132 & 90 & 3.4$\pm0.5$ & $2.1\times10^9$ & $8.4\times10^9$ \nl
PG 1119+120 & 200 & 0.050 & 260 & 4.5$\pm0.8$ & $4.2\times10^8$ &
$1.7\times10^9$ \nl
PG 1126-041 & 240 & \nodata & \nodata & $<2.6^b$ & $<3.4\times10^8$ & $<1.4\times10^9$ \nl
PG 1202+281 & 685 & \nodata & \nodata & $<2.4^b$ & $<2.3\times10^9$ & $<9.2\times10^9$ \nl
PG 1351+640 & 360 & 0.088 & 230 & 4.6$\pm1.0$ & $1.3\times10^9$ & $5.2\times10^9$ \nl
PG 1402+261 & 680 & \nodata & \nodata & $<1.3^b$ & $<1.2\times10^9$ & $<4.9\times10^9$ \nl
PG 1415+451 & 470 & 0.114 & 50 & 3.3$\pm0.6$ & $1.5\times10^9$ & $6.1\times10^9$ \nl
PG 1440+356 & 320 & 0.078 & 370 & 8.7$\pm1.4$ & $2.0\times10^9$ &
$8.0\times10^9$ \nl
PG 1613+658 & 530 & 0.129 & 490 & 8.5$\pm1.0$ & $5.0\times10^9$ &
$2.0\times10^{10}$ \nl
\enddata
\tablecomments{Also, for Mrk 1014, $L'_{\rm CO} = 10^{9.94}$
K km s$^{-1}$ pc$^2$, and for IZw1, $L'_{\rm CO} = 10^{9.70}$
K km s$^{-1}$ pc$^2$.}
\tablenotetext{a}{Luminosity Distance.}
\tablenotetext{b}{3$\sigma$ upper limits on $S_{\rm CO} \Delta v$ are
calculated assuming $\Delta v = 280$ km s$^{-1}$.}
\tablenotetext{c}{Calculated assuming $H_0 = 75$ km s$^{-1}$ Mpc$^{-1}$
and $q_0 = 0.5$.}
\tablenotetext{d}{Calculated assuming $\alpha = 4 M_{\odot}$ [K km s$^{-1}$
pc$^2]^{-1}$.}
\end{deluxetable}

\begin{figure}[h]
\plotfiddle{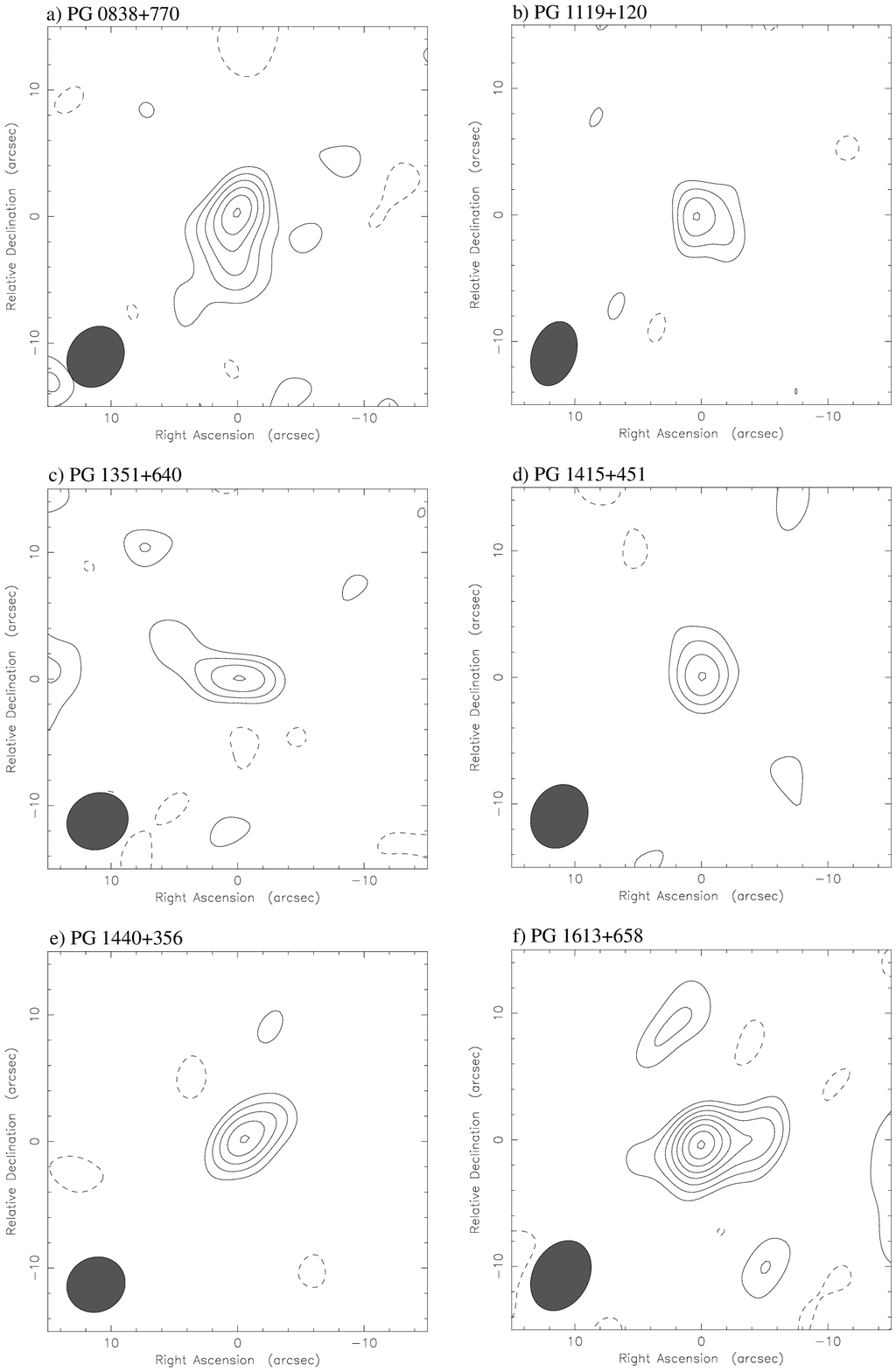}{9.0 in}{0}{90}{90}{-270}{-20}
\caption{}
\end{figure}

\begin{figure}[h]
\plotfiddle{Evans.fig2.ps}{9.0 in}{0}{90}{90}{-270}{0}
\caption{}
\end{figure}

\begin{figure}[h]
\plotfiddle{Evans.fig3.ps}{9.0 in}{0}{90}{90}{-270}{0}
\caption{}
\end{figure}

\begin{figure}[h]
\plotfiddle{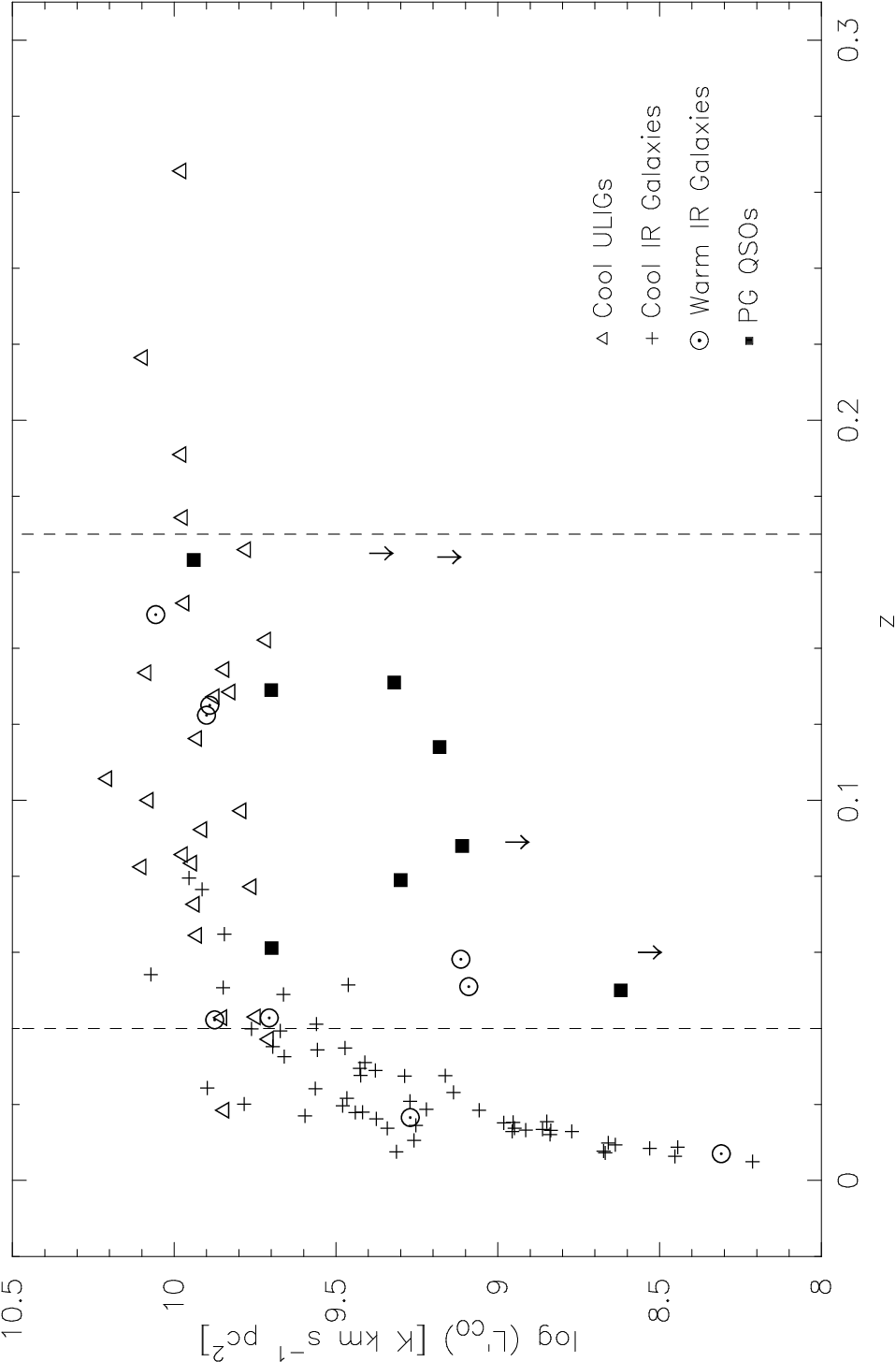}{9.0 in}{0}{90}{90}{-270}{0}
\caption{}
\end{figure}

\begin{figure}[h]
\plotfiddle{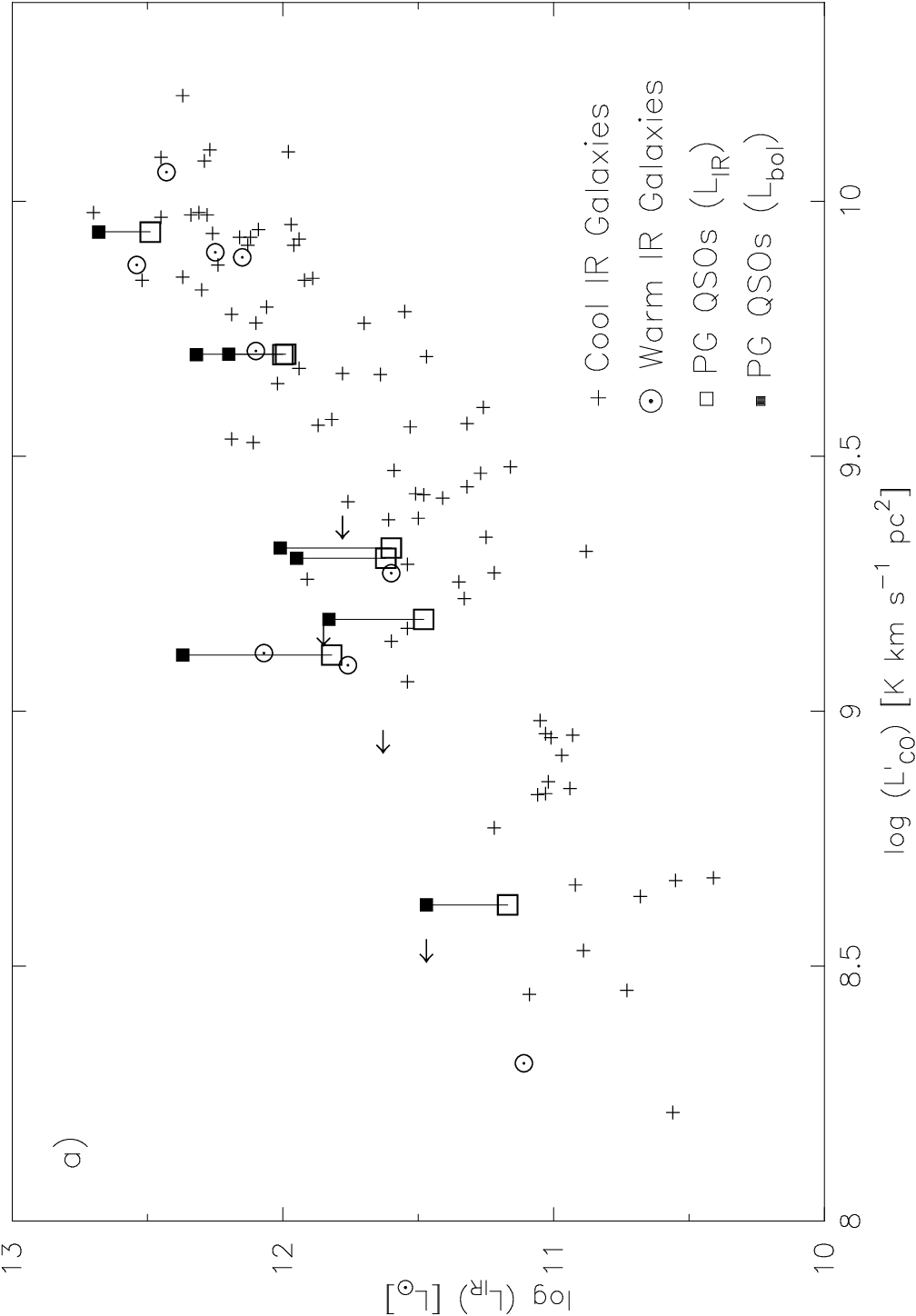}{9.0 in}{0}{90}{90}{-270}{0}
\figurenum{5a}
\caption{}
\end{figure}

\begin{figure}[h]
\plotfiddle{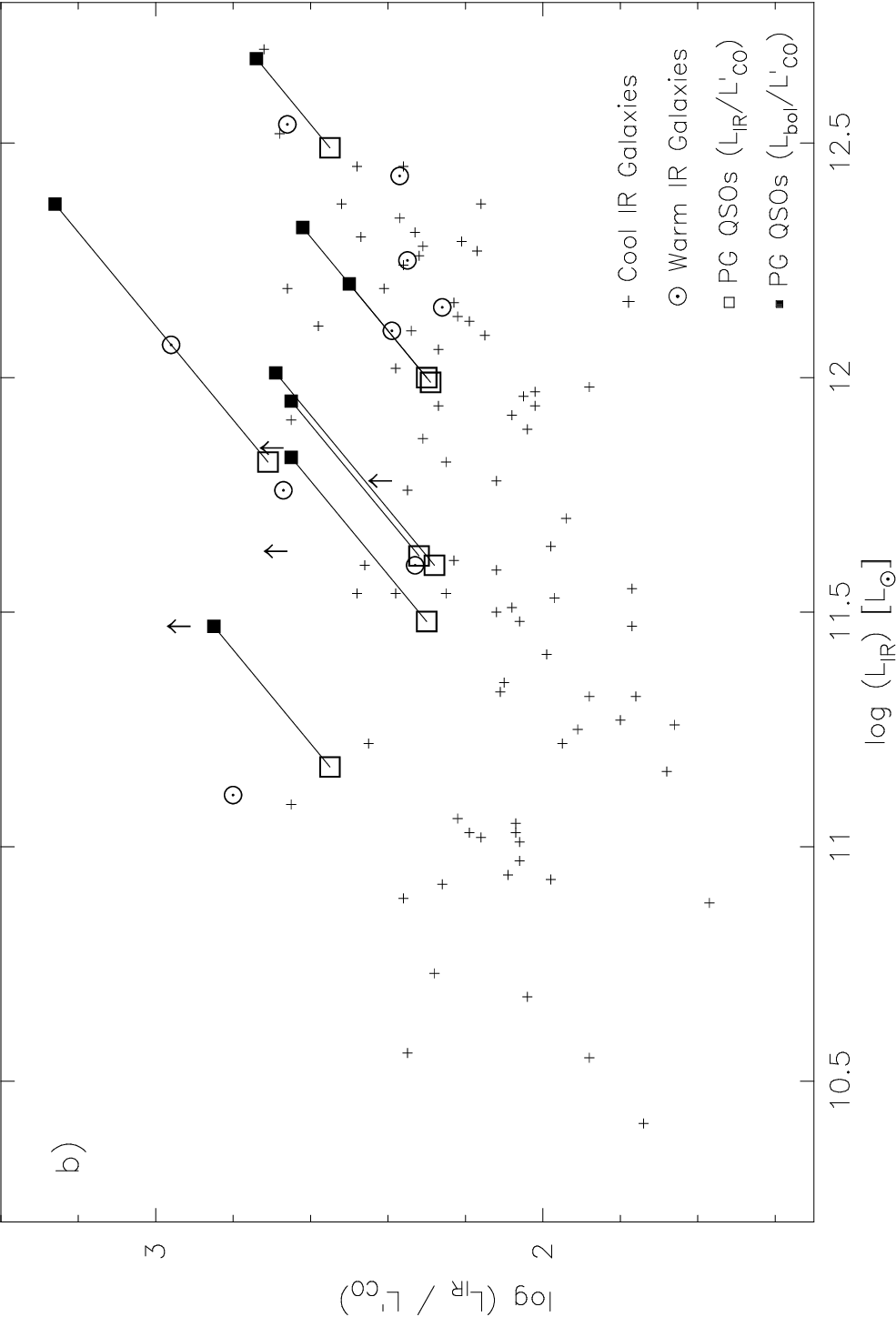}{9.0 in}{0}{90}{90}{-270}{0}
\figurenum{5b}
\caption{}
\end{figure}

\begin{figure}[h]
\plotfiddle{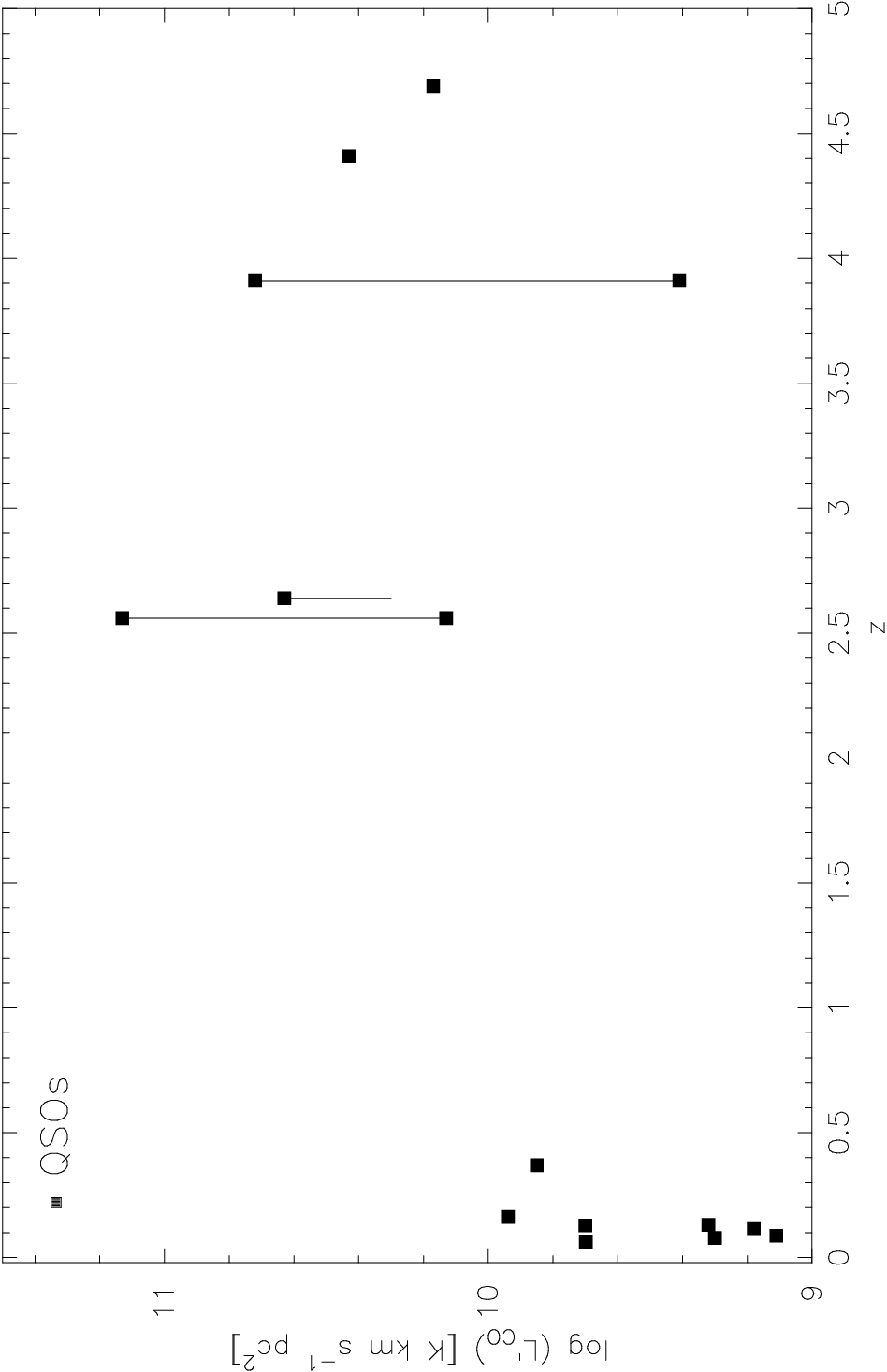}{9.0 in}{0}{90}{90}{-270}{0}
\figurenum{6}
\caption{}
\end{figure}


\begin{thebibliography}{}

\bibitem[]{}
Alloin, D., Barvainis, R., Gordon, M. A., \& Antonucci, R. R. J.
1992, A\&A, 265, 429

\bibitem[]{}
Bahcall, J. N., Kirhakos, S., Saxe, D. H., Schneider, D. P. 1997, 
ApJ, 479, 642

\bibitem[]{}
Barger, A. J. et al. 1998, Nature, 394, 248

\bibitem[]{}
Barger, A. J., Cowie, L. L., \& Sanders, D. B. 1999, ApJ, 518, 5

\bibitem[]{}
Barnes, J. E. \& Hernquist, L. 1996, ApJ, 471, 115

\bibitem[]{}
Barthel, P. D. 1989, ApJ, 336, 606

\bibitem[]{}
Barvainis, R., Alloin, D., \& Antonnuci, R. 1989, ApJ, 337, L69

\bibitem[]{}
Barvainis, R., Alloin, D., Guilloteau, S. \& Antonnuci, R. 1998, ApJ, 492,
L13

\bibitem[]{}
Barvainis, R., Tacconi, L., Antonucci, R., Alloin, D., \& Coleman, P.
1994, Nature, 371, 586

\bibitem[]{}
Bryant, P. M. \& Scoville, N. Z. 1999, AJ, 117, 2632

\bibitem[]{}
Downes, D., Neri, R., Wiklind, T., Wilner, D. J., \& Shaver, P.
1999, ApJ, 513, L1

\bibitem[]{}
Downes, D. \& Solomon, P. M. 1998, ApJ, 507, 615

\bibitem[]{}
Evans, A. S., Kim, D.-C., Mazzarella, J. M., Scoville, N. Z.,
\& Sanders, D. B. 1999, ApJ, 520, L107

\bibitem[]{}
Evans, A. S., Sanders, D. B., Cutri, R. M., Radford, S. J. E.,
Surace, J. A., Solomon, P. M., Downes, D., \& Kramer, C. 1998, ApJ,
506, 205

\bibitem[]{}
Evans, A. S., Sanders, D. B., Mazzarella, J. M., Solomon, P. M.,
Downes, D., Kramer, C., \& Radford, S. J. E. 1996, ApJ, 457, 658

\bibitem[]{}
Evans, A. S., Surace, J. A., Mazzarella, J. M., \& Sanders, D. B. 2000,
in preparation 

\bibitem[]{}
Ferrarese, L. \& Merritt, D. 2000, ApJ, 539, L9

\bibitem[]{}
Frayer, D. T., Ivison, R. J., Scoville, N. Z., Evans, A. S., Yun, M. S.,
Smail, I., Barger, A. J., Blain, A. W., \& Kneib, J.-P. 1999, ApJ, 514, L13

\bibitem[]{}
Gebhardt, K. et al. 2000, ApJ, 539, L13

\bibitem[]{}
Guilloteau, S., Omont, A., McMahon, R. G., Cox, P., \& Petitjean, P 1997,
A\&A, 328, L1

\bibitem[]{}
G\"{u}sten, R., Serabyn, E., Kasemann, C., Schinckel, A., Schneider, G.,
Schulz, A., \& Young, K. 1993, \apj, 402, 537

\bibitem[]{}
Hughes, D. H. et al. 1998, Nature, 394, 241

\bibitem[]{}
Kormendy, J., Bender, R., Evans, A. S., \& Richstone, D.  1998, AJ, 115,
1823

\bibitem[]{}
Kormendy, J. \& Sanders, D. B. 1992, ApJ, 390, L53

\bibitem[]{}
Magorrian, J. et al. 1998, AJ, 115, 2285

\bibitem[]{}
Ohta, K., Yamada, T., Nakanishi, K, Kohno, K., Akiyama, M.,
\& Kawabe, R. 1996, Nature, 382, 426

\bibitem[]{}
Omont, A., Petitjean, P., Guilloteau, S., McMahon, R. G., Solomon,
P. M., \& Pecontal, E. 1996, Nature, 382, 428

\bibitem[]{}
Radford, S. J. E., Solomon, P. M., \& Downes, D. 1991, ApJ, 368, L15

\bibitem[]{}
Sakamoto, K., Okumura, S. K., Ishizuki, S., \& Scoville, N. Z. 1999,
ApJ, 525, 691

\bibitem[]{}
Sanders, D. B., Scoville, N. Z., \& Soifer, B. T. 1988, ApJ, 335, L1

\bibitem[]{}
Sanders, D. B., Scoville, N. Z., \& Soifer, B. T. 1991, ApJ, 370, 158

\bibitem[]{}
Sanders, D. B., Scoville, N. Z., Tilanus, R. P. J., Wang, Z., \& Zhou, S.
1993, in Back to the Galaxy, eds S. Holt and F. Verter (New York: AIP),
311

\bibitem[]{}
Sanders, D. B., Scoville, N. Z., Zensus, A., Soifer, B. T., Wilson, T. L.,
Zylka, R., Steppe, H. 1989b, A\&A, 213, L5

\bibitem[]{}
Sanders, D. B., Phinney, E. S., Neugebauer, G., Soifer, B. T., \&
Matthews, K. 1989a, ApJ, 347, 29

\bibitem[]{}
Sanders, D.B., Soifer, B.T., Elias, J.H., Madore, B.F., Matthews, K.,
Neugebauer, G., \& Scoville, N.Z.  1988a, \apj , 325, 74

\bibitem[]{}
Sanders, D.B., Soifer, B.T., Elias, J.H., Neugebauer, G. \& Matthews, K.
1988b, \apj , 328, L35

\bibitem[]{}
Schinnerer, E., Eckart, A., \& Tacconi, L. J. 1998, ApJ, 500, 147

\bibitem[]{}
Schmidt, M. \& Green, R. F. 1993, ApJ, 269, 352

\bibitem[]{}
Scoville, N. Z., Carlstrom, J. C., Chandler, C. J., Phillips, J. A.,
Scott, S. L., Tilanus, R. P., \& Wang, Z. 1992, PASP, 105, 1482

\bibitem[]{}
Scoville, N. Z., Padin, S., Sanders, D. B., Soifer, B. T., \&
Yun, M. S. 1993, ApJ, 415, 75 

\bibitem[]{}
Scoville, N. Z. \& Sanders, D. B. 1987, in Interstellar Processes, ed. D.
Hollenbach \& H. Thronson (Dordrecht: Reidel), 21

\bibitem[]{}
Smail, I, Ivison, R. J., \& Blain, A. W. 1997, ApJL, 490, L5

\bibitem[]{}
Solomon, P. M., Radford, S. J. E., \& Downes, D. 1992,
Nature, 356, 318

\bibitem[]{}
Solomon, P. M., Downes, D., Radford, S. J. E., \& Barrett, J. W. 1997, ApJ,
478, 144

\bibitem[]{}
Shepherd, M. C., Pearson, T. J., \& Taylor, G. B. 1995, BAAS, 27, 903

\bibitem[]{}
Strong, A. W. et al. 1988, A\&A, 207, 1

\bibitem[]{}
Surace, J. A. 1998, PhD Thesis, University of Hawaii

\bibitem[]{}
Surace, J. A. \& Sanders, D. B. 2000, AJ, submitted (SS00) 

\bibitem[]{}
Surace, J. A., Sanders, D. B., Vacca, W. D., Veilleux, S., \& Mazzarella,
J. M. 1998, ApJ, 492, 116

\bibitem[]{}
Trentham, N. 2000, MNRAS, in press

\bibitem[]{}
van der Marel, R. P. 1999, AJ, 117, 774

\bibitem[]{}
Veilleux, S., Kim, D.-C., \& Sanders, D. B. 1999,
ApJ, 522, 113

\bibitem[]{}
Veilleux, S., Kim, D.-C., Sanders, D. B., Mazzarella,
J. M., \& Soifer, B. T. 1995, ApJS, 98, 171

\end{thebibliography}
\end{document}